\documentclass[final,authoryear,5p,times,twocolumn]{elsarticle}
\usepackage{amsmath}
\usepackage[usenames,dvipsnames]{color}
\usepackage{booktabs}
\journal{Journal of Quantitative Spectroscopy \& Radiative Transfer}

\newcommand{\tripoli}{{\sc Tripoli-4}\textsuperscript{ \textregistered}}

\begin{document}

\begin{frontmatter}

\title{Poisson-Box Sampling algorithms for three-dimensional Markov binary mixtures}


\author[label1]{Coline Larmier}
\address[label1]{Den-Service d'Etudes des R\'eacteurs et de Math\'ematiques Appliqu\'ees (SERMA), CEA, Universit\'e Paris-Saclay, 91191 Gif-sur-Yvette, FRANCE.}
\author[label1]{Andrea Zoia\corref{cor1}}
\cortext[cor1]{Corresponding author. Tel. +33 (0)1 69 08 79 76}
\ead{andrea.zoia@cea.fr}
\author[label1]{Fausto Malvagi}
\author[label2]{Eric Dumonteil}
\address[label2]{IRSN, 31 Avenue de la Division Leclerc, 92260 Fontenay aux Roses, FRANCE.}
\author[label1]{Alain Mazzolo}





\begin{abstract}
Particle transport in Markov mixtures can be addressed by the so-called Chord Length Sampling (CLS) methods, a family of Monte Carlo algorithms taking into account the effects of stochastic media on particle propagation by generating on-the-fly the material interfaces crossed by the random walkers during their trajectories. Such methods enable a significant reduction of computational resources as opposed to reference solutions obtained by solving the Boltzmann equation for a large number of realizations of random media. CLS solutions, which neglect correlations induced by the spatial disorder, are faster albeit approximate, and might thus show discrepancies with respect to reference solutions. In this work we propose a new family of algorithms (called 'Poisson Box Sampling', PBS) aimed at improving the accuracy of the CLS approach for transport in $d$-dimensional binary Markov mixtures. In order to probe the features of PBS methods, we will focus on three-dimensional Markov media and revisit the benchmark problem originally proposed by Adams, Larsen and Pomraning~\cite{benchmark_adams} and extended by Brantley~\cite{brantley_benchmark}: for these configurations we will compare reference solutions, standard CLS solutions and the new PBS solutions for scalar particle flux, transmission and reflection coefficients. PBS will be shown to perform better than CLS at the expense of a reasonable increase in computational time.
\end{abstract}


\end{frontmatter}

%\linenumbers

\section{Introduction}

Linear particle transport theory in random media is key to several applications in nuclear science and engineering, such as neutron diffusion in pebble-bed reactors or randomly mixed water-vapour phases in boiling water reactors~\cite{pomraning, larsen, levermore, sanchez, wong}, and inertial confinement fusion~\cite{zimmerman, zimmerman_adams, haran}. Material and life sciences as well as radiative transport also often involve particle propagation in random media~\cite{torquato, NatureOptical, davis, kostinski, clouds, tuchin, brantley_jeju}.

In this context, the material cross sections composing the traversed medium and the particle sources are distributed according to some statistical laws, and the physical observable of interest is typically the ensemble-averaged angular particle flux $\langle \varphi({\bf r}, {\boldsymbol \omega}) \rangle$, namely,
\begin{equation}
\langle \varphi({\bf r}, {\boldsymbol \omega}) \rangle = \int {\cal P}(q) \varphi^{(q)}({\bf r}, {\boldsymbol \omega}) dq,
\end{equation}
where $\varphi^{(q)}({\bf r}, {\boldsymbol \omega})$ satisfies the linear Boltzmann equation corresponding to a single realization $q$, and ${\cal P}(q)$ is the stationary probability of observing the state $q$ for the material cross sections and/or the sources~\cite{pomraning, renewal}. In the following, we consider linear particle transport in binary stochastic mixing composed of two immiscible random media (say $\alpha$ and $\beta$).

Exact solutions for $\langle \varphi \rangle $, or more generally for some ensemble-averaged functional $\langle F[ \varphi] \rangle$ of the particle flux, can be obtained using a so-called quenched disorder approach: an ensemble of medium realizations are first sampled from the underlying mixing statistics; then, the linear transport equation is solved for each realization by either deterministic or Monte Carlo methods, and the physical observables of interest $F[ \varphi]$ are determined; ensemble averages are finally computed. In a series of recent papers, we have provided reference solutions for particle transport in $d$-dimensional random media with Markov statistics~\cite{larmier_benchmark, larmier_models}, where the spatial disorder has been generated by means of homogeneous and isotropic $d$-dimensional Poisson tessellations~\cite{larmier}.

Reference solutions for particle transport in stochastic media are computationally expensive, so faster but approximate methods have been therefore proposed. A first approximate approach consists in deriving an expression for the ensemble-averaged flux $\langle \varphi \rangle$ in each material: this generally leads to an infinite hierarchy of equations, which ultimately requires a closure formula, such as in the celebrated Levermore-Pomraning model~\cite{pomraning, levermore, su}. A second approach is based on Monte Carlo algorithms that reproduce the ensemble-averaged solutions to various degrees of accuracy by modifying the displacement laws of the simulated particles in order to take into account the effects of spatial disorder~\cite{zimmerman_adams, sutton, donovan}. The Chord Length Sampling (CLS) algorithm is perhaps the most representative and best-known example of such algorithms: the basic idea behind CLS is that the interfaces between the constituents of the stochastic medium are sampled on-the-fly during the particle displacements by drawing the distances to the following material boundaries from a distribution depending on the mixing statistics. It has been shown that the CLS algorithm formally solves the Levermore-Pomraning model for Markovian binary mixing~\cite{zimmerman_adams, sahni1, sahni2}. The free parameters of the CLS model are the average chord length $\Lambda_i$ through each material, and the volume fraction $p_i$. Since the spatial configuration seen by each particle is regenerated at each particle flight, the CLS corresponds to an annealed disorder model, as opposed to the quenched disorder of the reference solutions, where the spatial configuration is frozen for all the traversing particles. This means that the correlations on particle trajectories induced by the spatial disorder are neglected in the standard implementation of CLS. Generalization of these Monte Carlo algorithms including partial memory effects due to correlations for particles crossing back and forth the same materials have been also proposed~\cite{zimmerman_adams}.

CLS, which had been originally formulated for Markov statistics, has been extensively applied also to randomly dispersed spherical inclusions into background matrices, with application to pebble-bed and very high temperature gas-cooled reactors~\cite{sutton, donovan}. In order to quantify the accuracy of CLS with respect to reference solutions for spherical inclusions, several comparisons have been proposed in two and three dimensions~\cite{sutton, donovan, brantley_martos, brantley_lp}. Some methods to mitigate the errors between CLS and the reference solutions have been presented in the context of eigenvalue calculations, e.g., in~\cite{ji_brown}. For Markov mixing specifically, a number of benchmark problems comparing CLS and reference solutions have been proposed in the literature so far~\cite{benchmark_adams, brantley_benchmark, renewal, brantley_conf, brantley_conf_2} with focus on $1d$-geometries (either of the rod or slab type); flat $2d$ geometries have been considered in~\cite{haran}. These benchmark comparisons have been recently extended to $d$-dimensional Markov geometries, for $d=2$ (extruded) and $d=3$~\cite{larmier_cls}.

Not surprisingly, CLS solutions may display discrepancies as compared to reference solutions, whose relevance varies strongly with the system dimensionality, the average chord length and the material volume fraction~\cite{larmier_cls}. For the case of $1d$ slab geometries with Markov mixing, possible improvements to the standard CLS algorithm accounting for partial memory effects for particle trajectories have been detailed~\cite{zimmerman_adams}, and numerical tests have revealed that these corrections contribute to palliating the discrepancies~\cite{brantley_benchmark}, although a generalization to higher dimensions seems hardly feasible with reasonable computational burden~\cite{zimmerman_adams}.

In this work we propose a new family of Monte Carlo algorithms aimed at improving the standard CLS for $d$-dimensional Markov media, yet keeping the increase in algorithmic complexity to a minimum. Inspiration comes from the observation that the physical observables related to particle transport through quasi-isotropic Poisson tessellations based on Cartesian boxes are almost identical to those computed for isotropic Poisson tessellations, for any dimension $d$~\cite{larmier_models, larmier_criticality}, which confirms the considerations in~\cite{mikhailov}. This quite remarkable property suggests that the standard CLS algorithm can be extended by replacing the memoryless sampling of material interfaces by the sampling of $d$-dimensional Cartesian boxes sharing the statistical features of quasi-isotropic Poisson tessellations, so as to mimic the spatial correlations that would be induced by isotropic Poisson tessellations. We will call this class of algorithms Poisson Box Sampling (PBS).

In order to illustrate the behaviour of the PBS with respect reference solutions and to CLS, we will revisit the classical benchmark problem for transport in Markov binary mixtures proposed by Adams, Larsen and Pomraning~\cite{benchmark_adams} and revisited by Brantley~\cite{brantley_benchmark}. The physical observables of interest will be the particle flux $\langle \varphi \rangle$, the transmission coefficient $\langle T \rangle$ and the reflection coefficient $\langle R \rangle$, for incident flux conditions and for uniform interior sources.

This paper is organized as follows: in Sec.~\ref{benchmark_definition} we will recall the benchmark specifications that will be used for our analysis in dimension $d=3$. In Sec.~\ref{quenched_approach} we will illustrate the reference solutions for the benchmark problem obtained by using isotropic and quasi-isotropic Poisson tessellations: this preliminary investigation will allow establishing that quasi-isotropic tessellations yield results very close to those of isotropic tessellations, as expected based on previous investigations. Then, in Sec.~\ref{annealed_approach} we will describe in detail the PBS algorithms, compare these methods to the reference solutions and to the standard CLS approach, and discuss their respective merits and drawbacks. Conclusions will be finally drawn in Sec.~\ref{conclusions}.

\section{Benchmark specifications}
\label{benchmark_definition}

In order for this paper to be self-contained, we briefly recall here the benchmark specifications that have been selected for this work, which are essentially drawn from those originally proposed in~\cite{benchmark_adams} and~\cite{renewal}, and later extended in~\cite{brantley_benchmark, brantley_conf, brantley_conf_2}.

We consider mono-kinetic linear particle transport through a stochastic binary medium with homogeneous and isotropic Markov mixing. The medium is non-multiplying, with isotropic scattering. The geometry consists of a cubic box of side $L=10$ (in arbitrary units), with reflective boundary conditions on all sides of the box except two opposite faces (say those perpendicular to the $x$ axis), where leakage boundary conditions are imposed. Two kinds of sources will be considered: either an imposed normalized incident angular flux on the leakage surface at $x=0$ (with zero interior sources), or a distributed homogeneous and isotropic normalized interior source (with zero incident angular flux on the leakage surfaces). The benchmark configurations pertaining to the former kind of source will be called {\em suite} I, whereas those pertaining to the latter will be called {\em suite} II~\cite{brantley_benchmark}. Markov mixing statistics are entirely defined by assigning the average chord length for each material $i = \alpha, \beta$, namely $\Lambda_i$. The (homogeneous) probability $p_i$ of finding material $i$ at an arbitrary location within the box follows from
\begin{equation}
p_{i}= \frac{\Lambda_i}{\Lambda_i + \Lambda_j}.
\end{equation}
By definition, the material probability $p_i$ yields the volume fraction for material $i$. The cross sections for each material will be denoted as customary $\Sigma_i$ for the total cross section and $\Sigma_{s,i}$ for the scattering cross section. The average number of particles surviving a collision in material $i$ will be denoted by $c_i = \Sigma_{s,i} / \Sigma_i \le 1$. The physical parameters for the benchmark configurations are recalled in Tabs.~\ref{tab_param1} and~\ref{tab_param2}: the benchmark specifications include three cases (numbered $1$, $2$ and $3$, corresponding to different materials), and three sub-cases (noted $a$, $b$ and $c$, corresponding to different $c_i $ for a given material) for each case~\cite{benchmark_adams}.

\begin{table}[!ht]
\begin{center}
\begin{tabular}{lcccc}
\toprule
Case & $\Sigma_\alpha$ & $\Lambda_\alpha$ & $\Sigma_\beta$ & $\Lambda_\beta$ \\
\midrule
1 & 10/99 & 99/100 & 100/11 & 11/100 \\
2 & 10/99 & 99/10 & 100/11 & 11/10 \\
3 & 2/101 & 101/20 & 200/101 & 101/20 \\
\bottomrule
\end{tabular}
\end{center}
\caption{Material parameters for the three cases of the benchmark configurations.}
\label{tab_param1}
\end{table}

\begin{table}[!ht]
\begin{center}
\begin{tabular}{lccc}
\toprule
Sub-case & a & b & c \\
\midrule
$c_\alpha$ & 0 & 1 & 0.9 \\
$c_\beta$ & 1 & 0 & 0.9 \\
\bottomrule
\end{tabular}
\end{center}
\caption{Scattering probabilities for the three sub-cases of the benchmark configurations.}
\label{tab_param2}
\end{table}

Following~\cite{brantley_benchmark}, the physical observables of interest for the benchmark will be the ensemble-averaged outgoing particle currents $\langle J \rangle $ on the two surfaces with leakage boundary conditions, the ensemble-averaged scalar particle flux $\langle \varphi(x) \rangle= \langle \int \int \int\varphi({\bf r}, {\boldsymbol \omega}) d {\boldsymbol \omega} dy dz \rangle$ along $0 \le x \le L$, and the total scalar flux $\langle \varphi \rangle =  \langle \int \varphi(x) dx \rangle$. For the {\em suite} I configurations, the outgoing particle current on the side opposite to the imposed current source represents the ensemble-averaged transmission coefficient, namely, $\langle T \rangle = \langle J_{x=L} \rangle $, whereas the outgoing particle current on the side of the current source represents the ensemble-averaged reflection coefficient, namely, $\langle R \rangle = \langle J_{x=0} \rangle $. For the {\em suite} II configurations, the outgoing currents on opposite faces are expected to be equal (within statistical fluctuations), for symmetry reasons. In this case, we also introduce the average leakage current $\langle L \rangle = \langle (T+R)/2 \rangle$.

\section{Reference solutions}
\label{quenched_approach}

In view of computing reference solutions for particle transport in three-dimensional quenched disorder, the generation of Markov mixing statistics will be based on random tessellations, which are stochastic aggregates of disjoint and space-filling cells obeying a given probability distribution~\cite{santalo}. In the following we will describe the two kinds of stochastic geometries that will be used for our analysis, namely isotropic Poisson tessellations and quasi-isotropic Poisson tessellations. Reference solutions for a $d$-dimensional generalization of the Adams, Larsen and Pomraning benchmark with homogeneous and isotropic Markov mixing have been thoroughly described in~\cite{larmier_benchmark}, where the ensemble-averaged scalar particle flux $\langle \varphi(x) \rangle$ and the currents $\langle R \rangle$ and $\langle T \rangle$ have been determined. In this section we recall the methods and the key results, and detail the changes and the additions that have been made with respect to our previous work.

\subsection{Isotropic Poisson tessellations}
\label{poisson}

Three-dimensional homogeneous and isotropic Poisson tessellations are obtained by partitioning an arbitrary domain with random planes sampled from an auxiliary Poisson process~\cite{santalo, miles1964, miles1972}. A single free parameter $\rho_{\cal P}$ (which is called the tessellation density) is required, which formally corresponds to the average number of planes of the tessellation that would be intersected by an arbitrary segment of unit length. An explicit construction amenable to Monte Carlo realizations for $2d$ geometries of finite size had been established in~\cite{switzer} (for a numerical investigation see, e.g.,~\cite{haran, lepage}), and recently generalized to $d$-dimensional domains~\cite{mikhailov}. The algorithm for $3d$ tessellations of a cube of side $L$ has been detailed in~\cite{larmier}. An example of realization of homogeneous and isotropic Poisson tessellation is provided in Fig.~\ref{geo_poisson}.

Isotropic Poisson geometries satisfy a Markov property: for domains of infinite size, arbitrary lines drawn through the tessellation will be cut by the surfaces of the polyhedra into segments whose lengths $\ell$ are exponentially distributed, with average chord length $\langle \ell \rangle = 1/\rho_{\cal P}$~\cite{santalo}. The quantity $\Lambda = \langle \ell \rangle$ intuitively defines the correlation length of the Poisson geometry, i.e, the typical linear size of a polyhedron composing the random tessellation~\cite{pomraning}.

\subsection{Colored stochastic geometries}

Homogeneous and isotropic binary Markov mixtures required for the reference solutions corresponding to the benchmark specifications are obtained as follows: first, an isotropic Poisson tessellation is constructed as described above. Then, each polyhedron of the geometry is assigned a material composition by formally attributing a distinct `color' $\alpha$ or $\beta$, with associated complementary probabilities $p_\alpha$ and $p_\beta = 1-p_\alpha$~\cite{pomraning}. This gives rise to (generally) non-convex $\alpha$ and $\beta$ clusters, each composed of a random number of convex polyhedra. An example of realization for a colored Poisson tessellation is shown in Fig.~\ref{geo_poisson}.

The average chord length $\Lambda_\alpha$ through clusters with composition $\alpha$ is related to the correlation length $\Lambda$ of the geometry via $\Lambda = (1-p_\alpha) \Lambda_\alpha$, and for $\Lambda_\beta$ we similarly have $\Lambda =  p_\alpha \Lambda_\beta$. This yields $1 / \Lambda_\alpha + 1 / \Lambda_\beta =  1 / \Lambda$, and we recover
\begin{equation}
p_\alpha = \frac{\Lambda}{\Lambda_\beta} = \frac{\Lambda_\alpha}{\Lambda_\alpha + \Lambda_\beta}.
\end{equation}
Based on the formulas above, and using $\rho_{\cal P} = 1/\Lambda$, the parameters of the colored Poisson geometries corresponding to the benchmark specifications provided in Tab.~\ref{tab_param1} are easily derived.

\subsection{Poisson-Box tessellations}
\label{box}

Box tessellations refer to a class of anisotropic stochastic geometries composed of Cartesian parallelepipeds with random sides~\cite{santalo}. The special case of Poisson-Box tessellations was proposed in~\cite{miles1972}: a domain is partitioned by randomly generated planes orthogonal to the three axes $x$, $y$ and $z$ through a Poisson process of intensity $\rho_x$, $\rho_y$ and $\rho_z$, respectively. We will assume that the three parameters are equal, namely, $\rho_x=\rho_y=\rho_z= \rho_{\cal B}$, which leads to homogeneous quasi-isotropic Poisson tessellations of density $\rho_{\cal B}$.

\begin{figure}[t]
\begin{center}
\,\,\,\, a) \,\,\,\,\\
\includegraphics[width=0.42\columnwidth]{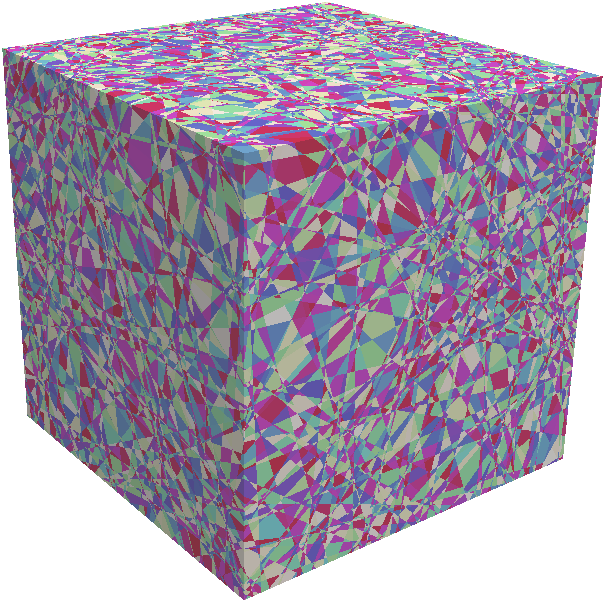}\,\,\,\,
\includegraphics[width=0.42\columnwidth]{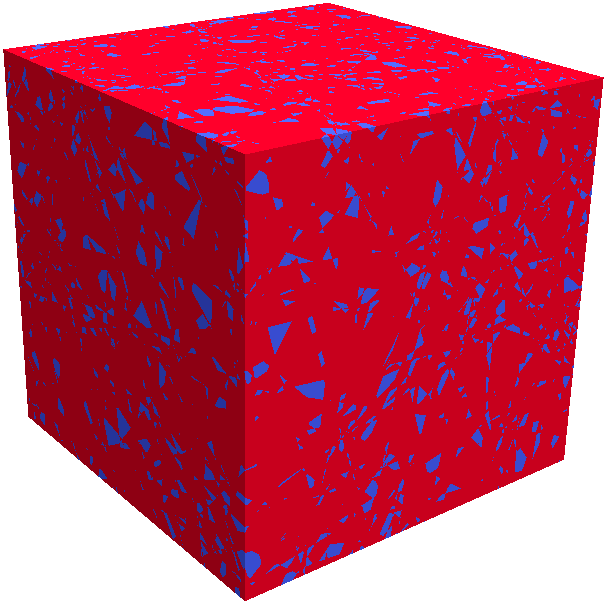}\\
\,\,\,\, b) \,\,\,\,\\
\includegraphics[width=0.42\columnwidth]{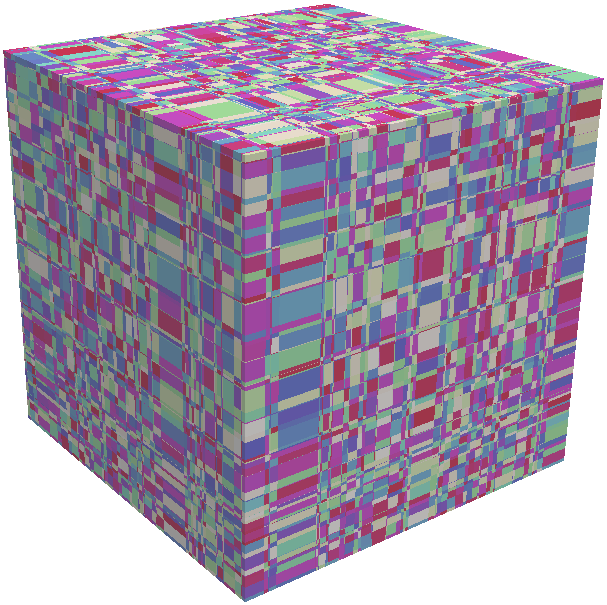}\,\,\,\,
\includegraphics[width=0.42\columnwidth]{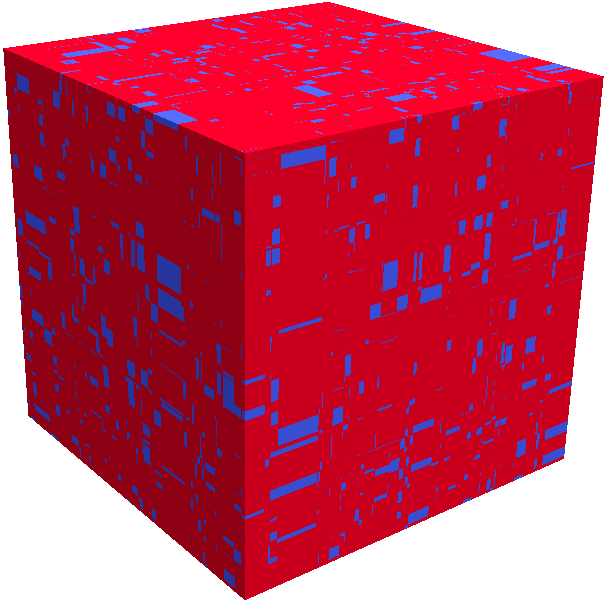}\\
\end{center}
\caption{Examples of realizations of a) homogeneous isotropic Poisson tessellations b) homogeneous Box tessellations, corresponding to the Case $1$ of the benchmark specifications ($\Lambda_{\alpha}=99/100$, $\Lambda_{\beta}=11/100$), before (left) and after (right) attributing the material label, with probability $p=0.9$ of assigning the label $\alpha$. Red corresponds to the label $\alpha$ and blue to the label $\beta$. The size of the cube is $L=10$.}
\label{geo_poisson}
\end{figure}

The explicit construction for Poisson-Box tessellations restricted to a cubic box of side $L$ has been provided in~\cite{larmier_models}. An example of realization of Poisson-Box tessellation is illustrated in Fig.~\ref{geo_poisson}. The chord distribution through Poisson-Box tessellations is not exponential; its average can be computed exactly, and yields $\Lambda = (2/3) / \rho_{\cal B}$~\cite{miles1972, larmier_models}. The coloring procedure is identical to that of isotropic Poisson tessellations, and the properties of the average chord lengths through colored clusters carry over as they stand: an example of colored realization is shown in Fig.~\ref{geo_poisson}. In order to avoid confusion with the Poisson tessellations described above, we will refer to Poisson-Box geometries simply as Box tessellations in the following.

\subsection{Comparing Poisson and Box tessellations}

In a series of benchmark calculations for multiplying and non-multiplying systems we have shown by Monte Carlo simulations that Box tessellations yield physical observables related to particle transport that are very close to those computed for isotropic Poisson tessellations~\cite{larmier_models, larmier_criticality}, which confirms the findings in~\cite{mikhailov}. Since this property represents a crucial step towards the construction of the PBS algorithms that will be presented in Sec.~\ref{annealed_approach}, we would like to preliminarily verify that this peculiar feature carries over to the Adam, Larsen and Pomraning benchmark configurations.

The key point is that both tessellations depend on a single free parameter, namely the average chord length $ \Lambda$, in addition to the coloring probability $p$. For isotropic Markovian binary mixtures, the average chord length of the Poisson tessellation and the coloring probability are chosen so that the resulting average chord lengths in the colored clusters, namely $ \Lambda_{\alpha}$ and $ \Lambda_{\beta}$, match the correlation lengths in the random media~\cite{pomraning}.

A natural choice is therefore to set $ \Lambda_{\alpha}$ and $ \Lambda_{\beta}$ to be equal for Poisson and Box tessellations, which ensures that the two colored geometries are `statistically equivalent'. As shown in~\cite{larmier_models}, this can be achieved by choosing the same parameters $ \Lambda$ and $p$ for the two tessellations. Correspondingly, we have a constraint on the tessellation densities $\rho_{\cal P}$ and $\rho_{\cal B}$, which must now satisfy
\begin{equation}
\frac{1}{ \Lambda} =\rho_{\cal P}=\frac{3}{2} \rho_{\cal B}.
\label{calib}
\end{equation}
Numerical simulations show that by imposing Eq.~\eqref{calib} the chord length distributions for the two tessellations are barely distinguishable~\cite{larmier_models, larmier_criticality}, which is of utmost importance since the properties of particle transport through random media mostly depend on the shape of the chord length distribution~\cite{pomraning}.

The similarity of the chord length distributions is all the more striking when considering that other geometrical features do not share comparable affinities. For instance, if we set the average chord length $\Lambda$ to be equal for the two tessellations as in Eq.~\eqref{calib}, the average volumes of a typical polyhedron read
\begin{align}
& \langle V \rangle_{\cal P} = \frac{6}{\pi} \Lambda^3\\
& \langle V \rangle_{\cal B} = \left( \frac{3}{2}\right)^3 \Lambda^3,
\label{volume_stat}
\end{align}
respectively, i.e., the average volume of the Box tessellations is much larger than that of the Poisson tessellations~\cite{larmier_models}.
 
\subsection{Simulation results for reference solutions}
\label{simulation_results_quenched}

For each benchmark configuration, a large number $M$ of geometries has been generated, and the material properties have been attributed to each volume as described in~\cite{larmier_benchmark}. Then, for each realization $k$ of the ensemble, linear particle transport has been simulated by using the production Monte Carlo code \tripoli{}, developed at CEA~\cite{T4}. \tripoli{} is a general-purpose stochastic transport code capable of simulating the propagation of neutral and charged particles with continuous-energy cross sections in arbitrary geometries. In order to comply with the benchmark specifications, constant cross sections adapted to mono-energetic transport and isotropic angular scattering have been prepared. The number of simulated particle histories per configuration is $10^6$. For a given physical observable ${\cal O}$, the benchmark solution is obtained as the ensemble average
\begin{equation}
\langle {\cal O} \rangle = \frac{1}{M} \sum_{k=1}^M {\cal O}_k,
\end{equation}
where ${\cal O}_k$ is the Monte Carlo estimate for the observable ${\cal O}$ obtained for the $k$-th realization. Specifically, currents $R_k$ and $T_k$ at a given surface are estimated by summing the statistical weights of the particles crossing that surface. Scalar fluxes $\varphi_k(x)$ have been tallied using the standard track length estimator over a pre-defined spatial grid containing $10^2$ uniformly spaced meshes along the $x$ axis.

The error affecting the average observable $\langle {\cal O} \rangle$ results from two separate contributions, the dispersion
\begin{equation}
\sigma^2_G = \frac{1}{M} \sum_{k=1}^M {{\cal O}_k}^2 - {\langle {\cal O} \rangle}^2
\end{equation}
of the observables exclusively due to the stochastic nature of the geometries and of the material compositions, and 
\begin{equation}
\sigma^2_{{\cal O}}=\frac{1}{M} \sum_{k=1}^M \sigma_{{\cal O}_k}^2,
\end{equation}
which is an estimate of the variance due to the stochastic nature of the Monte Carlo method for particle transport, $\sigma_{{\cal O}_k}^2$ being the dispersion of a single calculation~\cite{donovan, sutton}. The statistical error on $\langle {\cal O} \rangle$ is then estimated as
\begin{equation}
\sigma[ \langle {\cal O}\rangle ] = \sqrt{\frac{\sigma^2_G}{M}+\sigma^2_{{\cal O}}}.
\end{equation}

The reference solutions corresponding to isotropic Poisson tessellations have been first presented in~\cite{larmier_benchmark} with $M=10^3$ realizations for every benchmark configuration. In order to reduce the dispersion of the observables due to the statistical nature of the geometries, a new set of reference solutions has been computed in~\cite{larmier_cls} by increasing the number of realizations for the benchmark configurations displaying larger correlation lengths (i.e., larger material chunks). The data for reference solutions presented here are taken from~\cite{larmier_cls}: we have set $M=2 \times 10^4$ for the sub-case $2b$ of the \textit{suite} II; $M=5 \times 10^3$ for all the other sub-cases of case $2$; $M=5 \times 10^4$ for the sub-case $3b$ of the \textit{suite} II; and $M=10^4$ for all the other sub-cases of case $3$. For all remaining cases and sub-cases, we have used the same number of realizations as in~\cite{larmier_benchmark}, namely, $M=10^3$. Additionally, reference solutions corresponding to Box tessellations have been computed for each benchmark configuration by following the same procedure as above, and the number of realizations has been set equal to that of the corresponding Poisson tessellations. 

Particle transport calculations have been run on a cluster based at CEA, with Intel Xeon E5-2680 V2 2.8 GHz processors. For the simulations discussed here considerable speed-ups have been obtained for the most fragmented geometries thanks to the possibility of reading pre-computed connectivity maps for the volumes composing the geometry, which largely increases the performances of particle tracking. 

Reference solutions for both tessellations are provided in Tabs.~\ref{tab_suite1_case1} to \ref{tab_suite1_case3} for the benchmark cases corresponding to {\em suite} I, and in Tabs.~\ref{tab_suite2_case1} to \ref{tab_suite2_case3} for the benchmark cases corresponding to {\em suite} II, respectively: the ensemble-averaged total scalar flux $\langle \varphi \rangle$, transmission coefficient $\langle T \rangle$, and reflection coefficient $\langle R \rangle$ are displayed for Poisson and Box tessellations. The respective computer times are also provided in the same tables. The ensemble-averaged spatial flux $\langle \varphi(x) \rangle$ is illustrated in Figs.~\ref{fig_space_1} to \ref{fig_space_3}. As mentioned above, the reference solutions for Poisson geometries are taken from reference~\cite{larmier_cls}; the reference solutions for Box tessellations have never been presented before.

Simulation results for the Adam, Larsen and Pomraning benchmark configurations basically confirm our previous findings: the physical observables related to particle transport through Box tessellations are very close to those of isotropic Poisson tessellations, which was expected based on their respective chord length distributions being very similar. The agreement between the two sets of results increases by decreasing the average chord length (i.e., for more fragmented tessellations). An exception must be remarked for sub-case $1b$ of {\em suite} I, in particular for the transmission coefficient $\langle T \rangle$, despite this configuration being highly fragmented. Since this sub-case is composed of absorbing chunks dispersed in a scattering background, the observed discrepancy might be attributed to the effects induced by the shape of the chunks on particle transport (which are different for the two tessellations, as noticed above). For the spatial flux profiles, slight differences emerge for the less fragmented configurations, e.g., sub-cases $2b$, $3a$ and $3b$ of {\em suite} I.

The computer time required for the reference solutions (as shown in Tabs.~\ref{tab_suite1_case1} to \ref{tab_suite1_case3}) depends on the material compositions and increases with the complexity of the configurations, i.e., with the number of polyhedra composing the tessellation. For a given average chord length $\Lambda$, the average number of volumes is smaller in Box geometries than in Poisson geometries, which follows from the expressions of the typical volumes in Eq.~\eqref{volume_stat}. Transport simulations in Box tessellations are faster than in Poisson tessellations for configurations composed of a large number of polyhedra, such as those of case $1$; for cases $2$ and $3$, finite-size effects due to $\Lambda$ being comparable to $L$ come into play, and computer times become almost identical for Poisson and Box geometries.

\section{Approximate solutions: from CLS to PBS}
\label{annealed_approach}

Reference solutions based on the quenched disorder approach are very demanding in terms of computational resources, so that intensive research efforts have been devoted to the development of Monte Carlo-based annealed disorder models capable of approximating the effects of spatial disorder on-the-fly during particle trajectories, i.e., within a single transport simulation. In this section we first briefly recall the standard CLS algorithm, for the sake of completeness, and then introduce a new class of Monte Carlo methods, called Poisson Box Sampling (PBS), combining the principles of CLS with the generation of material volumes inspired by the findings concerning Box tessellations. Simulation results of the PBS for the benchmark configurations will be compared to those of CLS and to the reference solutions obtained above.

\subsection{Chord Length Sampling (CLS)}

The annealed disorder algorithms initially developed by Zimmerman and Adams go now under the name of Chord Length Sampling methods~\cite{zimmerman, zimmerman_adams}. The standard form of CLS (Algorithm A in~\cite{zimmerman_adams}) formally solves the Levermore-Pomraning equations corresponding to Markov mixing with the approximation that memory of the crossed material interfaces is lost at each particle flight~\cite{sahni1, sahni2}. Algorithm A has the following structure~\cite{zimmerman_adams}: 
\begin{itemize}
\item Step $1$: each particle history begins by sampling position, angle and velocity from the specified source, as customary. Moreover, the particle is assigned a supplementary attribute, the material label, which is sampled according to the volume fraction probability $p_i$.
\item Step $2$: three distances are computed: the distance $\ell_b$ to the next physical boundary, along the current direction of the particle; the distance $\ell_c$ to collision, which is determined by using the material cross section chosen at the previous step: if the particle has a material label $\alpha$, e.g., then $\ell_c$ will be drawn from an exponential distribution of parameter $1/\Sigma_\alpha$; and the distance $\ell_i$ to material interface, which is sampled from an exponential distribution with parameter $\Lambda_\alpha$, i.e., the average chord length of material $\alpha$, if the particle has a material label $\alpha$ (whence the name of CLS).
\item Step $3$: the minimum distance among $\ell_b, \ell_c$ and $\ell_i$ has to be selected. If the minimum is $\ell_b$, the particle is moved along a straight line until the external boundary is hit (the direction is updated in the case of reflection); if the minimum is $\ell_c$, the particle is moved to the collision point, and the outgoing particle features are selected according to the collision kernel pertaining to the current material label; if the minimum is $\ell_i$, the particle is moved to the interface between the two materials, and the material label is switched. If the particle has not been absorbed, return to Step $2$.
\end{itemize}

The particle will ultimately either get absorbed in one of the chunks or leak out of the boundaries of the random medium. As observed above, Algorithm A assumes that the particle has no memory of its past history, and in particular the crossed interfaces are immediately forgotten (which is consistent with the closure formula of the Levermore-Pomraning model). In this respect, CLS is an approximation of the exact treatment of disorder-induced spatial correlations. In particular, CLS is expected to be less accurate in the presence of strong scatterers with optically thick mean material chunk length~\cite{brantley_benchmark, larmier_cls}. A thorough discussion of the shortcomings of the CLS approach for $d=1$ can be found, e.g., in~\cite{ji_cls}.

\subsection{Poisson Box Sampling (PBS)}

For the case of $1d$ slab geometries, two improved versions of CLS Algorithm A have been proposed in the literature, by partially taking into account the memory effects induced by the spatial correlations~\cite{zimmerman_adams}: in the former, called Algorithm B, instead of sampling the material interfaces one at a time a full random slab is generated, and particles do not switch material properties until either the forward or the backward surfaces of the slab are crossed; in the latter, called Algorithm C, a slab is generated as in Algorithm B, and the slab traversed before entering the current one is also kept in memory. The basic idea behind Algorithms B and C is to preserve the shape and the position of the material chunks (thus partially restoring spatial correlations) by generating an additional typical random slab whenever particles cross the material surfaces of the current volume.

As expected, Algorithms B and C have been shown to approximate the reference solutions for Markov mixing in $1d$ more accurately than Algorithm A, at the expense of an increased computational cost~\cite{brantley_benchmark, zimmerman_adams}. Algorithm B in particular has been extensively tested for the Adams, Larsen and Pomraning benchmark in slab geometries, and performs better than Algorithm A for all configurations~\cite{brantley_benchmark}. As observed in~\cite{zimmerman_adams}, it is not trivial to extend Algorithms B and C to higher dimensions: this can be immediately understood by remarking that randomly generating a typical material chunk in dimension three with Markov mixing would correspond to sampling a typical polyhedral cell of the isotropic Poisson tessellations, whose exact distributions for the volume, surface, number of faces, etc., are unfortunately unknown to this day~\cite{miles1972, santalo, larmier}. In dimension one the typical chunk is a slab of exponentially distributed width, which considerably simplifies the computational burden.

A possible way to overcome this issue and improve Algorithm A in higher dimensions is however suggested by the numerical findings concerning Box tessellations. Since the chord length distribution of Box tessellations is very close to that of Poisson tessellations, it seems reasonable to extend Algorithm B by generating on-the-fly the typical cells of Box tessellations, i.e., Cartesian boxes with exponentially distributed side lengths. The generalization of Algorithm C would immediately follow by keeping memory of the last visited box. We will call this new class of Monte Carlo algorithms Poisson Box Sampling (PBS), and we will denote by PBS-1 the former (inspired by Algorithm B) and by PBS the latter (inspired by Algorithm C). In view of the aforementioned similarity between quasi-isotropic and isotropic Poisson tessellations, intuitively we expect that PBS methods will preserve the increased accuracy of Algorithms B and C over Algorithm A, yet allowing for a relatively straightforward construction and a fairly minor additional computational burden.

By adapting the strategy of CLS, the algorithm for PBS-1 proceeds as follows:
\begin{itemize}
\item Step $1$: initialize each particle history by sampling position, angle and velocity from the specified source. In addition, a random Cartesian box is generated. The box is defined by its material label $i$ and its spatial position, given by the coordinates $(x_c,y_c,z_c)$ of its center and its sides: $l_x$, $l_y$ and $l_z$. Three pairs of random numbers, namely, $(\Delta_x^+,\Delta_x^-)$, $(\Delta_y^+,\Delta_y^-)$ and $(\Delta_z^+,\Delta_z^-)$, are sampled from independent exponential distributions of average $3\Lambda/2$. For {\em suite} I, we do not sample the value of $\Delta_x^-$, and we set $\Delta_x^-=0$. Then, we set the center of the box
\begin{align}
& x_c=x+(\Delta_x^+ - \Delta_x^-)/2,\\
& y_c=y+(\Delta_y^+ - \Delta_y^-)/2,\\
& z_c=z+(\Delta_z^+ - \Delta_z^-)/2,
\end{align}
and the sides
\begin{align}
& l_x=\Delta_x^+ + \Delta_x^-,\\
& l_y=\Delta_y^+ + \Delta_x^-,\\
& l_z=\Delta_z^+ + \Delta_z^-.
\end{align}
The material label of the box is sampled according to $p_i$.
\item Step $2$: we compute three distances: the distance $\ell_b$ to the next physical boundary, along the current direction of the particle; the distance $\ell_c$ to collision, which is determined by using the material cross section that has been chosen at the previous step: if the particle is in a box with material label $\alpha$, e.g., then $\ell_c$ will be drawn from an exponential distribution of parameter $1/\Sigma_\alpha$; and the distance $\ell_i$ to the next interface of the current box along the particle direction (the boundaries of the box being easily determined).
\item Step $3$: the minimum distance among $\ell_b, \ell_c$ and $\ell_i$ has to be selected: if the minimum is $\ell_b$, the particle is moved along a straight line until the external boundary is hit (the direction is updated in the case of reflection); if the minimum is $\ell_c$, the particle is moved to the collision point, and the outgoing particle features are selected according to the collision kernel pertaining to the current material label; if the minimum is $\ell_i$, the particle is moved along a straight line until the interface of the current box is hit: a new box is sampled as detailed below, and the new box becomes the current box. If the particle has not undergone a capture, return to Step $2$.
\end{itemize}

For the sampling of a new box at Step $3$, we begin by drawing a random spacing $\delta$ from an exponential distribution with average $3\Lambda/2$. Without loss of generality, if the interface of the current box hit by the particle is perpendicular to the $x$-axis, we set the following values for the side $l_x$ of the new box and the position $x_c$ of its center: $l_x=\delta$, $x_c=x+ l_x \, \omega_x /|\omega_x|$, where $\omega_x$ is the particle direction along the $x$-axis. The other features of the current box, namely, $l_y$, $l_z$, $y_c$ and $z_c$, are left unchanged for the new box (as suggested by the construction of Box tessellations). We would proceed in the same way for the $y$- and $z$-axis. Finally, the label of the new box is randomly sampled according to the coloring probability $p_i$.

Contrary to Algorithm A, the correlations induced by spatial disorder are partially preserved by the PBS-1 algorithm: indeed, each particle will see the same material properties until the current box is left. Moreover, when a new box is created, its features strongly depend on those of the previous box. This should globally improve the accuracy of PBS-1 with respect to CLS in reproducing the reference solutions for the benchmark. Long-range correlations spanning more than a box (i.e., a linear size of the order of $\Lambda$) are nonetheless suppressed, so that we still expect some discrepancies between PBS-1 solutions and those obtained by the quenched disorder approach for either Poisson or Box tessellations.

In order to further improve the accuracy of the PBS methods, we propose a second method, inspired by Algorithm C, that will be denoted PBS-2. The strategy is exactly as in the PBS-1 algorithm, the only difference being in the fact that, once a new box has been sampled, the old box is not deleted but is kept in memory (size, position and material label) until a new material interface is selected. If the particle leaves the new box by another interface, the old box is definitively deleted, another box is sampled and the new box becomes the old box. If the selected interface is the one that has been kept in memory, the new box will simply be the old box, and the roles are reversed. This implementation intuitively extends the range of spatial correlations, and is thus supposed to correspondingly enhance the accuracy with respect to reference solutions, at the expense of increasing the computational burden, too.

\subsection{Simulation results}
\label{simulation_results_cls}

The simulation results corresponding to CLS and PBS for the total scalar flux $\langle \varphi \rangle$, the transmission coefficient $\langle T \rangle$ and the reflection coefficient $\langle R \rangle$ are provided in Tabs.~\ref{tab_suite1_case1} to \ref{tab_suite1_case3} for the benchmark cases corresponding to {\em suite} I, and in Tabs.~\ref{tab_suite2_case1} to \ref{tab_suite2_case3} for the benchmark cases corresponding to {\em suite} II, respectively. The spatial flux $\langle \varphi(x) \rangle$ is illustrated in Figs.~\ref{fig_space_1} to \ref{fig_space_3}. For the CLS and PBS simulations of the benchmark configurations we have used 10$^9$ particles (10$^3$ replicas with 10$^6$ particles per replica), with resulting statistical uncertainties associated to each physical observable ${\cal O}$ denoted by $\sigma_\text{CLS}[ {\cal O}]$ and $\sigma_\text{PBS}[ {\cal O}]$, respectively.

Generally speaking, the solutions computed with PBS-1 show a better agreement with respect to the reference solutions based on Poisson tessellations than those computed with CLS, and overall remarkably well approximate the benchmark observables. Moreover, as expected from the previous considerations, PBS-2 shows a further enhanced accuracy with respect to PBS-1. A single exception has been detected for sub-case $1b$ of {\em suite} I, as reported in Tab.~\ref{tab_suite1_case1} and in Fig.~\ref{fig_space_1}. For this configuration, the results of the Box tessellations are slightly different from those of Poisson tessellations, as observed above,  for the spatial flux and the transmission coefficient. It turns out that both PBS algorithms provide results that are in excellent agreement with the reference solutions for the Box tessellation, which is consistent with their implementation. However, because of the observed discrepancy between Box and Poisson tessellations for sub-case $1b$, PBS show a small bias with respect to Poisson reference solutions. For the same case, CLS displays a better accuracy as compared to Poisson solutions, and this is most probably due to the fact that this algorithm exactly preserves isotropy.

The analysis of the approximate solutions suggests that the accuracy of CLS globally improves when decreasing the average chord length $\Lambda$: configurations pertaining to case $1$ globally show a better agreement than those of case $2$, and those of case $2$ show a better agreement than those of case $3$, as pointed out in~\cite{larmier_cls}. The improved PBS methods are less sensitive to the average chord length $\Lambda$ and show a satisfactory agreement for all benchmark configurations.

Computer times for the CLS and PBS solutions are also provided in Tabs.~\ref{tab_suite1_case1} to \ref{tab_suite1_case3}: not surprisingly, the approaches based on annealed disorder are much faster than the reference methods, since a single transport simulation is needed. PBS methods, while still much faster than reference solutions, for most configurations take sensibly longer than CLS: this is partly due to the increased complexity of the algorithms, and partly due to the fact that CLS is based on the sampling of the colored chord lengths (corresponding to clusters of polyhedra sharing all the same material label), whereas PBS require the sampling of un-colored boxes one at a time. Nonetheless, keeping in memory a further box amounts to an almost negligible additional computational burden for PBS-2 as opposed to PBS-1.

\section{Conclusions}
\label{conclusions}

In this paper we have proposed a new family of Monte Carlo methods aimed at approximating ensemble-averaged observables for particle transport in Markov binary mixtures, where reference results are obtained by sampling medium realizations from homogeneous and isotropic Poisson tessellations. The so-called Algorithm A of Chord Length Sampling method is perhaps the most widely adopted simulation tool to provide such approximate solutions, based on the Levermore-Pomraning model. Several numerical investigations have shown that Algorithm A works reasonably well in most cases, yet discrepancies between CLS and reference solutions may appear due to the fact that Algorithm A neglects the correlations induced by spatial disorder. For the case of one-dimensional slab geometries, two variants of the standard CLS method, namely Algorithm B and Algorithm C, have been proposed by partially including spatial correlations and memory effects. These algorithms provide an increased accuracy with respect to Algorithm A thanks to the on-the-fly generation of typical slabs during the particle displacements, but their generalization to higher dimensions appears to be non-trivial. A rigorous generalization in dimension three would for instance demand sampling on-the-fly typical polyhedra from homogeneous and isotropic Poisson tessellations, whose exact statistical distribution are unfortunately unknown.

In order to overcome these issues and derive CLS-like methods capable of taking into account spatial correlations for $d$-dimensional configurations, we have resorted to the key observation that quasi-isotropic Poisson tessellations (also called Box tessellations) based on Cartesian boxes yield chord length distributions and transport-related physical observables that in most cases are barely distinguishable from those coming from isotropic Poisson tessellations. This remarkable feature has inspired a generalization of CLS Algorithms B and C based on sampling on-the-fly random boxes obeying the same statistical properties as for Box tessellations. We have called these family of algorithms Poisson Box Sampling, or PBS.

We have proposed two variants of PBS: in PBS-1 we generate random $d$-dimensional boxes, similarly as done in Algorithm B of CLS, and in PBS-2 we additionally keep memory of the last generated box, in full analogy with Algorithm C of CLS. In order to test the performances of these new methods, we have compared PBS simulation results to the reference solutions and CLS solutions for the classical benchmark problem proposed by Adams, Larsen and Pomraning for particle propagation in stochastic media with binary Markov mixing. In particular, we have examined the evolution of the transmission coefficient, the reflection coefficient and the particle flux for the benchmark configurations in dimension $d=3$.

A preliminary investigation has shown that Poisson and Box tessellations lead to very similar results for all the benchmark configurations, as expected on the basis of previous works, which substantiates our motivation for PBS methods. Overall, the PBS-1 algorithm reproduces reference solutions based on Poisson tessellations more accurately that Algorithm A of CLS, at the expense of an increased computational cost. PBS-2 further increases the accuracy of PBS-1 by including memory effects and thus enhancing the range of spatial correlations that are correctly captured by the algorithm; the additional computational burden required by PBS-2 is almost negligible.

A local realization preserving (LRP) algorithm that extends the standard CLS in a way similar to PBS (i.e., by preserving information about the shape of the traversed polyhedra) has been independently developed at LLNL and tested against reference solutions and CLS Algorithm A~\cite{brantley_lrp}: in the future, it will be interesting to compare PBS to LRP. Moreover, future research work will be aimed at testing the performances of PBS methods as applied to other benchmark configurations with Markov mixtures, such as diffusing matrices with void or absorbing chunks~\cite{larmier_models}, or multiplying systems~\cite{larmier_criticality}.

\section*{Acknowledgements}
TRIPOLI-4\textsuperscript{ \textregistered} is a registered trademark of CEA. C.~Larmier, A.~Zoia, F.~Malvagi and A.~Mazzolo wish to thank \'Electricit\'e de France (EDF) for partial financial support.

\clearpage

\begin{table*}
\footnotesize
\begin{center}
\begin{tabular}{cccccccccc}
\toprule
Case & Algorithm & $\langle R \rangle$ & $\langle T \rangle$ & $\langle \varphi \rangle$ & $t_{\mathrm{tot}}$ [s] \\
\midrule
1a & Poisson & $0.4091 \pm 5\times 10^{-4}$ & $0.0163 \pm 10^{-4}$ & $6.328 \pm 0.007$ & $3.9 \times 10^{6}$\\
\cmidrule(lr){2-6}
 & Box & $0.4092	\pm 6\times 10^{-4}$ & $0.0166	\pm 10^{-4}$ & $6.321	\pm 0.008$ & $8.5 \times 10^{5}$\\
\cmidrule(lr){2-6}
 & CLS & $0.40176 \pm 2\times 10^{-5}$ &	$0.017491 \pm 4\times 10^{-6}$ & $6.3933 \pm 2\times 10^{-4}$ & $4.6 \times 10^{3}$\\
 & Err [$\%$] & $-1.79 \pm	0.13$ & $7.53 \pm	0.86$ & $1.03 \pm	0.12$\\
 \cmidrule(lr){2-6}
 & PBS-1 & $0.40683	\pm 2\times 10^{-5} $	& $0.017030	\pm 4\times 10^{-6} $ &	$6.3440	\pm 2\times 10^{-4}$ & 	$1.9\times10^{4}$ \\
 & Err [$\%$] & $-0.55 \pm	0.13$ & $4.70 \pm	0.84$ & $0.25 \pm	0.12$\\
  \cmidrule(lr){2-6}
 & PBS-2 & $0.40760	\pm 2\times 10^{-5} $ & 	$0.016898	\pm 4\times 10^{-6} $ & 	$6.3368	\pm 2\times 10^{-4}$ & 	$1.9\times10^{4}$ \\
 & Err [$\%$] & $-0.36 \pm 0.13$ & $3.88 \pm	0.83$ & $0.14 \pm	0.12$\\
\midrule
1b & Poisson & $0.0377 \pm 2\times 10^{-4}$ & $0.00085 \pm 3\times 10^{-5}$ & $1.918 \pm 0.003$ & $1.8 \times 10^{6}$ \\
\cmidrule(lr){2-6}
 & Box & $0.0379	\pm 2\times 10^{-4}$ & $0.00102	\pm 3\times 10^{-5}$ & $1.925	\pm 0.004$ & $3.4 \times 10^{5}$\\
\cmidrule(lr){2-6}
& CLS & $0.036714 \pm 6\times 10^{-6}$	& $0.0008413 \pm 9\times 10^{-7}$ & $1.91440 \pm 6\times 10^{-5}$ & $1.0 \times 10^{3}$ \\
 & Err [$\%$] & $-2.52 \pm	0.52$ &	$-1.03	\pm 3.46$	& $-0.20 \pm	0.17$ \\
  \cmidrule(lr){2-6}
 & PBS-1 & $0.036729	\pm 6\times 10^{-6} $ &	$0.001025 \pm	1\times 10^{-6}$ &	$1.91635	\pm 6\times 10^{-5} $ &	$5.8 \times 10^3$ \\
 & Err [$\%$] & $-2.48 \pm 0.52$ & $20.58 \pm	4.21$ & $-0.10 \pm 0.17$ \\
  \cmidrule(lr){2-6}
 & PBS-2 & $0.037188	\pm 6\times 10^{-6} $ &	$0.001028	\pm  10^{-6} $ &	$1.92029	\pm 6\times 10^{-5} $ &	$6.0\times 10^{3}$ \\
 & Err [$\%$] & $-1.26 \pm	0.52$ & $20.88 \pm	4.23$ & $0.11 \pm	0.17$\\
\midrule
1c &  Poisson & $0.4059 \pm 5\times 10^{-4}$ & $0.0164 \pm 10^{-4}$ & $6.303 \pm 0.008$ & $4.4 \times 10^{6}$\\
\cmidrule(lr){2-6}
 & Box & $0.4062	\pm 5\times 10^{-4}$ & $0.0168	\pm 10^{-4}$  &	$6.306 \pm	0.009$ & $8.5 \times 10^{5}$ \\
\cmidrule(lr){2-6}
& CLS & $0.39619 \pm 10^{-5}$ &	$0.016992 \pm 2\times 10^{-6}$ & $6.2957	\pm 2\times 10^{-4}$ & $1.1 \times 10^{4}$ \\
& Err [$\%$] & $-2.40 \pm	0.12$ & $3.62 \pm	0.84$ & $-0.12 \pm	0.13$ \\
 \cmidrule(lr){2-6}
 & PBS-1 & $0.40278 \pm	10^{-5} $ &	$0.017054	\pm 3\times 10^{-6} $ &	$6.3049 \pm	2\times 10^{-4}$ &	$4.7\times 10^{4}$ \\
 & Err [$\%$] & $-0.78 \pm	0.12$ & $4.00 \pm	0.85$ & $0.03 \pm	0.13$ \\
  \cmidrule(lr){2-6}
 & PBS-2 & $0.40399 \pm 10^{-5} $ &	$0.016998	\pm 3\times 10^{-6} $ &	$6.3082 \pm	2\times 10^{-4}$ &	$4.8\times 10^{4}$ \\
  & Err [$\%$] & $-0.48 \pm	0.12$ & $3.66 \pm	0.85$ & $0.08 \pm	0.13$\\
\bottomrule
\end{tabular}
\end{center}
\caption{Ensemble-averaged observables and computer time $t_{\mathrm{tot}}$ for the benchmark configurations: {\em suite} I - case $1$.\label{tab_suite1_case1}}
\end{table*}

\clearpage

\begin{table*}
\footnotesize
\begin{center}
\begin{tabular}{ccccccccccc}
\toprule
Case & Algorithm & $\langle R \rangle$ & $\langle T \rangle$ & $\langle \varphi \rangle$ & $t_{\mathrm{tot}}$ [s] \\
\midrule
2a &  Poisson & $0.225 \pm 0.001$ & $0.0937	\pm 4\times 10^{-4}$ & $7.57	\pm 0.01$ & $4.4 \times 10^{5}$\\
\cmidrule(lr){2-6}
 & Box & $0.228	\pm 0.001$ &	$0.0950	\pm 4\times 10^{-4}$ &	$7.54	\pm 0.01$ & $4.3\times 10^5$\\
 \cmidrule(lr){2-6}
 & CLS &  $0.20043 \pm 10^{-5}$ & $0.105624	\pm 9\times 10^{-6}$ & $7.6615 \pm 2\times 10^{-4}$ & $3.1 \times 10^{3}$\\
  & Err [$\%$] & $-11.08 \pm 0.45$ & $12.74 \pm	0.54$ & $1.22 \pm	0.13$\\
   \cmidrule(lr){2-6}
 & PBS-1 & $0.22066	\pm 10^{-5}$ & $0.098160	\pm 9\times 10^{-6}$ & $7.5601	\pm 2\times 10^{-4}$  &	$5.3 \times 10^{3}$\\
 & Err [$\%$] & $-2.11 \pm 0.50$ & $4.77 \pm	0.50$ & 	$-0.12 \pm	0.13$\\
  \cmidrule(lr){2-6}
 & PBS-2 & $0.22365 \pm	10^{-5}$ &	$0.097014	\pm 9\times 10^{-6}$ &	$7.5504	\pm 2\times 10^{-4}$ &	$5.3 \times 10^{3}$ \\
 & Err [$\%$] & $-0.79 \pm	0.51$ & $3.55 \pm	0.49$ & $-0.25 \pm	0.13$\\
\midrule
2b &  Poisson & $0.1616	\pm 8\times 10^{-4}$ & $0.1194	\pm 9\times 10^{-4}$ & $7.77	\pm 0.03$ & $3.4 \times 10^{5}$\\
\cmidrule(lr){2-6}
 & Box & $0.1626	\pm 9\times 10^{-4}$ & $0.1202	\pm 9\times 10^{-4}$ & $7.77 \pm	0.03$ & $2.9\times 10^5$\\
 \cmidrule(lr){2-6}
& CLS & $0.14223 \pm 10^{-5}$ & $0.10996 \pm 10^{-5}$ & $7.2609 \pm 2\times 10^{-4}$ & $9.3 \times 10^{2}$\\
 & Err [$\%$] & $-11.99 \pm	0.44$ & $-7.91 \pm	0.68$ & $-6.50 \pm 0.37$\\
  \cmidrule(lr){2-6}
 & PBS-1 & $0.14394	\pm 10^{-5}$ & $0.11168	\pm 10^{-5}$	& $7.3065	\pm 2\times 10^{-4}$ &	$3.1 \times 10^{3}$ \\
 & Err [$\%$] & $-10.94 \pm	0.45$ & $-6.48 \pm 	0.70$ & $-5.92 \pm 0.38$ \\
  \cmidrule(lr){2-6}
 & PBS-2 & $0.15193	\pm 10^{-5}$ & $0.11572	\pm 10^{-5}$ & $7.5152	\pm 2\times 10^{-4}$ &	 $3.2 \times 10^{3}$\\
 & Err [$\%$] & $-5.99 \pm	0.47$ & $-3.09 \pm	0.72$ & $-3.23 \pm	0.39$\\
\midrule
2c &  Poisson & $0.3457	\pm 5\times 10^{-4}$ & $0.1651	\pm 9\times 10^{-4}$ & $10.76	\pm 0.03$ & $4.8 \times 10^{5}$\\
\cmidrule(lr){2-6}
 & Box & $0.3474	\pm 5\times 10^{-4}$ & $0.1656	\pm 9\times 10^{-4}$ & $10.74	\pm 0.03$ & $4.0 \times 10^{5}$\\
 \cmidrule(lr){2-6}
& CLS & $0.27693 \pm 10^{-5}$ &	$0.15031 \pm 10^{-5}$	& $9.6048	\pm 2\times 10^{-4}$ & $8.9 \times 10^{3}$\\
 & Err [$\%$] & $-19.89 \pm	0.12$	& $-8.98 \pm	0.49$ &	$-10.73 \pm	0.23$ \\
  \cmidrule(lr){2-6}
 & PBS-1 & $0.32610	\pm 10^{-5}$ &	$0.16385	\pm 10^{-5}$	& $10.4372	\pm 2\times 10^{-4}$ &	$1.0 \times 10^{4}$ \\
 & Err [$\%$] & $-5.67 \pm	0.14$ & $-0.78 \pm 0.53$ & $-3.00 \pm	0.25$\\
  \cmidrule(lr){2-6}
 & PBS-2 & $0.33558	\pm 10^{-5}$ &	$0.16586	\pm 10^{-5}$ &	$10.6010	\pm 2\times 10^{-4}$	& $9.9 \times 10^{3}$ \\
 & Err [$\%$] & $-2.93 \pm	0.14$ &	$0.44 \pm	0.54$	& $-1.48 \pm	0.25$\\
\bottomrule
\end{tabular}
\end{center}
\caption{Ensemble-averaged observables and computer time $t_{\mathrm{tot}}$ for the benchmark configurations: {\em suite} I - case $2$.\label{tab_suite1_case2}}
\end{table*}

\clearpage

\begin{table*}
\footnotesize
\begin{center}
\begin{tabular}{ccccccccccc}
\toprule
Case & Algorithm & $\langle R \rangle$ & $\langle T \rangle$ & $\langle \varphi \rangle$ & $t_{\mathrm{tot}}$ [s] \\
\midrule
3a &  Poisson & $0.675	\pm 0.001$ & $0.1692	\pm 9\times 10^{-4}$ & $16.38	\pm 0.03$ & $1.4 \times 10^{6}$\\
  \cmidrule(lr){2-6}
 & Box & $0.677 \pm	0.001$ & $0.168 \pm	0.001$ & $16.39	\pm 0.03$ & $1.3 \times 10^6$\\
 \cmidrule(lr){2-6}
& CLS & $0.64107 \pm 2\times 10^{-5}$	& $0.19957	\pm 10^{-5}$ & $16.3231 \pm 6\times 10^{-4}$ & $9.1 \times 10^{3}$\\
& Err [$\%$] & $-5.06 \pm	0.20$ & $17.96 \pm	0.65$ & $-0.36 \pm	0.19$\\
  \cmidrule(lr){2-6}
 & PBS-1 & $0.64806	\pm 2\times 10^{-5}$ &	$0.19408	\pm 10^{-5}$ &	$16.3182	\pm 6\times 10^{-4}$ &	$8.1 \times 10^{3}$ \\
 & Err [$\%$] & $-4.02 \pm	0.21$ & $14.72 \pm	0.64$ & $-0.39 \pm	0.19$\\
  \cmidrule(lr){2-6}
 & PBS-2 & $0.66148	\pm 2\times 10^{-5}$ & $0.18226	\pm 10^{-5}$ & $16.3416	\pm 7\times 10^{-4}$ & $8.1 \times 10^{3}$ \\
 & Err [$\%$] & $-2.04 \pm	0.21$ & $7.73 \pm	0.60$ & $-0.25 \pm	0.19$ \\
\midrule
3b & Poisson & $0.0165	\pm 2\times 10^{-4}$ & $0.0457	\pm 9\times 10^{-4}$ & $3.47	\pm 0.03$ & $5.0 \times 10^{5}$\\
  \cmidrule(lr){2-6}
 & Box & $0.0166	\pm 2\times 10^{-4}$ & $0.0462	\pm 9\times 10^{-4}$ & $3.44	\pm 0.03$ & $4.1 \times 10^5$\\
 \cmidrule(lr){2-6}
& CLS & $0.012454 \pm 3\times 10^{-6}$ & $0.040345 \pm 6\times 10^{-6}$ & $3.2382 \pm 10^{-4}$ & $8.0 \times 10^{2}$ \\
& Err [$\%$] & $-24.48 \pm	0.97$ & $-11.80 \pm	1.68$ & $-6.55 \pm	0.70$\\
  \cmidrule(lr){2-6}
 & PBS-1 & $0.013974	\pm 4\times 10^{-6}$ & $0.044533	\pm 6\times 10^{-6}$ & $3.3178	\pm 10^{-4}$ & $1.2 \times 10^3$ \\
 & Err [$\%$] & $-15.26 \pm	1.09$	& $-2.65 \pm	1.86$ & $-4.26 \pm	0.71$\\
  \cmidrule(lr){2-6}
 & PBS-2 & $0.015823	\pm 4\times 10^{-6}$ & $0.047299	\pm 7\times 10^{-6}$ & $3.4216	\pm 2\times 10^{-4}$ & $1.2 \times 10^3$ \\
 & Err [$\%$] & $-4.05 \pm	1.23$ & $3.40 \pm	1.97$ & $-1.26 \pm	0.74$\\
\midrule
3c & Poisson & $0.3979 \pm	7\times 10^{-4}$ & $0.086	\pm 0.001$ & $7.89 \pm 0.03$ & $7.0 \times 10^{5}$\\
  \cmidrule(lr){2-6}
 & Box & $0.4008	\pm	7\times 10^{-4}$ & $0.086	\pm 0.001$ & $7.86	\pm 0.04$ & $6.9 \times 10^5$\\
 \cmidrule(lr){2-6}
& CLS & $0.34652 \pm 10^{-5}$ &	$0.080613	\pm 7\times 10^{-6}$ & $7.3217	\pm 2\times 10^{-4}$ & $8.8 \times 10^{3}$ \\
& Err [$\%$] & $-12.92 \pm	0.15$ & $-6.16 \pm 1.19$ & $-7.17 \pm	0.40$\\
  \cmidrule(lr){2-6}
 & PBS-1 & $0.36242 \pm	10^{-5}$ &	$0.086946	\pm 8\times 10^{-6}$ & $7.5230	\pm 2\times 10^{-4}$ & $7.6 \times 10^{3}$ \\
 & Err [$\%$] & $-8.92 \pm	0.15$ & $1.21 \pm	1.29$ & $-4.62 \pm	0.41$\\
  \cmidrule(lr){2-6}
 & PBS-2 & $0.38437	\pm 10^{-5}$ & $0.089701	\pm 8\times 10^{-6}$ & $7.7650	\pm 2\times 10^{-4}$ & $7.2 \times 10^3$ \\
 & Err [$\%$] & $-3.41 \pm	0.16$ & $4.42 \pm	1.33$ & $-1.55 \pm	0.42$\\
\bottomrule
\end{tabular}
\end{center}
\caption{Ensemble-averaged observables and computer time $t_{\mathrm{tot}}$ for the benchmark configurations: {\em suite} I - case $3$.\label{tab_suite1_case3}}
\end{table*}

\clearpage

\begin{table*}
\footnotesize
\begin{center}
\begin{tabular}{ccccccccccc}
\toprule
Case & Algorithm & $\langle L \rangle$ & $\langle \varphi \rangle$ & $t_{\mathrm{tot}}$ [s] \\
\midrule
1a & Poisson & $0.1583 \pm 3\times 10^{-4}$ & $7.530 \pm 0.008$ & $7.9 \times 10^{7}$\\
\cmidrule(lr){2-6}
 & Box & $0.1580 \pm	3\times 10^{-4}$ &	$7.533 \pm 0.008$ & $3.6 \times 10^{7}$\\
 \cmidrule(lr){2-6}
& CLS & $0.159828  \pm 8\times 10^{-6}$ & $7.4924 \pm 2\times 10^{-4}$ & $5.3 \times 10^{3}$\\
& Err [$\%$] & $0.98 \pm	0.17$ & $-0.49 \pm	0.10$ \\
  \cmidrule(lr){2-6}
 & PBS-1 & $0.158605	\pm 8\times 10^{-6}$ & $7.5218	\pm 2\times 10^{-4}$ & $2.3\times 10^4$\\
 & Err [$\%$] & $0.21 \pm	0.17$ & $-0.10 \pm 0.10$ \\
  \cmidrule(lr){2-6}
 & PBS-2 & $0.158416	\pm 8\times 10^{-6}$ & $7.5259	\pm 2\times 10^{-4}$ & $2.3\times 10^4$\\
 & Err [$\%$] & $0.09 \pm	0.17$ & $-0.05 \pm	0.10$ \\
\midrule
1b & Poisson & $0.0481 \pm 2\times 10^{-4}$ & $1.808	\pm 0.003$ & $7.4 \times 10^{7}$\\
\cmidrule(lr){2-6}
 & Box & $0.0481	\pm 2\times 10^{-4}$ & $1.820	\pm 0.003$ & $3.3 \times 10^{7}$\\
 \cmidrule(lr){2-6}
& CLS & $0.047859 \pm 5\times 10^{-6}$ & $1.79609	\pm 6\times 10^{-5}$ & $1.0 \times 10^{3}$\\
& Err [$\%$] & $-0.42 \pm	0.33$ & $-0.63 \pm	0.14$ \\
  \cmidrule(lr){2-6}
 & PBS-1 & $0.047910	\pm 5\times 10^{-6}$ & $1.80671	\pm 6\times 10^{-5}$ &	$5.5 \times 10^{3}$\\
 & Err [$\%$] & $-0.32 \pm	0.33$ & $-0.05	\pm 0.14$\\
  \cmidrule(lr){2-6}
 & PBS-2 & $0.048013	\pm 5\times 10^{-6}$ & $1.81201 \pm	6\times 10^{-5}$ &	$5.6 \times 10^{3}$\\
 & Err [$\%$] & $-0.10 \pm	0.33$ & $0.25 \pm	0.14$\\
\midrule
1c & Poisson & $0.1577 \pm 3\times 10^{-4}$ & $7.455 \pm 0.008$ & $7.7 \times 10^{7}$\\
\cmidrule(lr){2-6}
 & Box & $0.1576	\pm 3\times 10^{-4}$ & $7.470 \pm 0.008$ & $3.9 \times 10^{7}$ & \\
 \cmidrule(lr){2-6}
& CLS & $0.157383 \pm 6\times 10^{-6}$ & $7.3335	\pm 10^{-4}$ & $1.4 \times 10^{4}$\\
& Err [$\%$] & $-0.19 \pm	0.17$ & $-1.63 \pm	0.10$ \\
  \cmidrule(lr){2-6}
 & PBS-1 & $0.157630	\pm 6\times 10^{-6}$ & $7.4260	\pm 10^{-4}$ & $6.1 \times 10^{4}$\\
 & Err [$\%$] & $-0.04	\pm 0.17$ & $-0.39 \pm	0.11$ \\
  \cmidrule(lr){2-6}
 & PBS-2 & $0.157705	\pm 6\times 10^{-6}$ & $7.4434	\pm 10^{-4}$ & $6.3 \times 10^{4}$\\
 & Err [$\%$] & $0.01 \pm	0.17$ & $-0.15 \pm	0.11$\\
\bottomrule
\end{tabular}
\end{center}
\caption{Ensemble-averaged observables and computer time $t_{\mathrm{tot}}$ for the benchmark configurations: {\em suite} II - case $1$.\label{tab_suite2_case1}}
\end{table*}

\clearpage

\begin{table*}
\footnotesize
\begin{center}
\begin{tabular}{ccccccccccc}
\toprule
Case & Algorithm & $\langle L \rangle$ & $\langle \varphi \rangle$  & $t_{\mathrm{tot}}$ [s] \\
\midrule
2a & Poisson & $0.1892 \pm 3\times 10^{-4}$ & $7.27 \pm 0.01$ & $5.8 \times 10^{5}$\\
\cmidrule(lr){2-6}
 & Box & $0.1882	\pm 3\times 10^{-4}$ & $7.31	\pm 0.01$ & $4.4\times 10^5$\\
 \cmidrule(lr){2-6}
& CLS & $0.191527 \pm 9\times 10^{-6}$ & $6.8774	\pm 2\times 10^{-4}$ & $3.0 \times 10^{3}$\\
& Err [$\%$] & $1.21 \pm	0.16$ & $-5.36 \pm	0.18$ \\
  \cmidrule(lr){2-6}
 & PBS-1 & $0.189005	\pm 9\times 10^{-6}$ & $7.1825	\pm 2\times 10^{-4}$ & $5.7\times 10^{3}$\\
 & Err [$\%$] & $-0.12 \pm	0.15$ & $-1.16 \pm	0.19$\\
  \cmidrule(lr){2-6}
 & PBS-2 & $0.188756	\pm 9\times 10^{-6}$ & $7.2490	\pm 2\times 10^{-4}$ & $5.9\times 10^{3}$\\
 & Err [$\%$] & $-0.26 \pm	0.15$ & $-0.25 \pm	0.19$\\
\midrule
2b & Poisson & $0.1931  \pm 4\times 10^{-4}$ & $6.53 \pm	0.01$ & $1.7 \times 10^{6}$\\
\cmidrule(lr){2-6}
 & Box & $0.1939	\pm 4\times 10^{-4}$ & $6.63 \pm	0.01$ & $9.1\times 10^5$\\
 \cmidrule(lr){2-6}
& CLS & $0.181518	\pm 9\times 10^{-6}$ & $6.0577 \pm	2\times 10^{-4}$ & $8.6 \times 10^{2}$\\
& Err [$\%$] & $-6.01 \pm	0.21$ & $-7.31 \pm	0.19$\\
  \cmidrule(lr){2-6}
 & PBS-1 & $0.182655 \pm 9\times 10^{-6}$ & $6.1625	\pm 2\times 10^{-4}$ & $2.8 \times 10^3$\\
 & Err [$\%$] & $-5.42 \pm	0.21$ & $-5.70 \pm	0.19$\\
  \cmidrule(lr){2-6}
 & PBS-2 & $0.187874 \pm 9\times 10^{-6}$ & $6.3690 \pm	2\times 10^{-4}$ & $2.9 \times 10^3$\\
 & Err [$\%$] & $-2.72 \pm	0.21$ & $-2.54 \pm	0.20$ \\
\midrule
2c & Poisson & $0.2688 \pm 6\times 10^{-4}$ & $9.55	\pm 0.02$ & $4.9 \times 10^{5}$ \\
\cmidrule(lr){2-6}
 & Box & $0.2680	\pm 6\times 10^{-4}$ & $9.62	\pm 0.02$ & $3.3 \times 10^{5}$ &\\
 \cmidrule(lr){2-6}
& CLS &  $0.240117	\pm 8\times 10^{-6}$ & $8.3498 \pm	2\times 10^{-4}$ & $8.4 \times 10^{3}$\\
& Err [$\%$] & $-10.69 \pm	0.20$ & $-12.58 \pm	0.18$\\
\cmidrule(lr){2-6}
 & PBS-1 & $0.260937	\pm 9\times 10^{-6}$ & $9.2561	\pm	2\times 10^{-4}$ & $9.8 \times 10^3$ \\
 & Err [$\%$] & $-2.94 \pm	0.22$ & $-3.09 \pm	0.20$\\
  \cmidrule(lr){2-6}
 & PBS-2 & $0.265038	\pm 9\times 10^{-6}$ & $9.4484	\pm	2\times 10^{-4}$ & $9.7 \times 10^3$\\
 & Err [$\%$] & $-1.42 \pm	0.22$ & $-1.08 \pm	0.20$\\
\bottomrule
\end{tabular}
\end{center}
\caption{Ensemble-averaged observables and computer time $t_{\mathrm{tot}}$ for the benchmark configurations: {\em suite} II - case $2$.\label{tab_suite2_case2}}
\end{table*}

\clearpage

\begin{table*}
\footnotesize
\begin{center}
\begin{tabular}{ccccccccccc}
\toprule
Case & Algorithm & $\langle L \rangle$ & $\langle \varphi \rangle$  & $t_{\mathrm{tot}}$ [s] \\
\midrule
3a &  Poisson &  $0.4098 \pm 4\times 10^{-4}$ & $22.82	\pm 0.07$ & $1.7 \times 10^{6}$\\
\cmidrule(lr){2-6}
 & Box & $0.4088 \pm 4\times 10^{-4}$ & $23.33	\pm 0.08$ & $1.6 \times 10^6$ & \\
 \cmidrule(lr){2-6}
& CLS &  $0.40807 \pm 10^{-5}$ & $19.7173 \pm	6\times 10^{-4}$  & $1.1 \times 10^{4}$\\
& Err [$\%$] & $-0.42 \pm	0.10$ & $-13.61 \pm	0.28$ \\
\cmidrule(lr){2-6}
 & PBS-1 & $0.40795	\pm 10^{-5}$ & $20.7792	\pm 7\times 10^{-4}$ & $1.0 \times 10^{4}$ \\
 & Err [$\%$] & $-0.45 \pm	0.10$ & $-8.96 \pm	0.29$\\
  \cmidrule(lr){2-6}
 & PBS-2 & $0.40856	\pm 10^{-5}$ & $22.0417	\pm 8\times 10^{-4}$ &	$1.1 \times 10^{4}$ \\
 & Err [$\%$] & $-0.30 \pm	0.10$ &	$-3.43 \pm	0.31$\\
\midrule
3b & Poisson &  $0.0868	\pm 3\times 10^{-4}$ & $2.98	\pm 0.01$ & $8.7 \times 10^{5}$ \\
\cmidrule(lr){2-6}
 & Box & $0.0864	\pm 3\times 10^{-4}$ & $3.02 \pm	0.01$ & $7.2 \times 10^5$\\
 \cmidrule(lr){2-6}
& CLS &  $0.080949 \pm	6\times 10^{-6}$ & $2.6747 \pm 10^{-4}$ & $7.8 \times 10^{2}$\\
& Err [$\%$] & $-6.78 \pm 0.34$ & $-10.12 \pm	0.30$\\
\cmidrule(lr){2-6}
 & PBS-1 & $0.082950	\pm 6\times 10^{-6}$ & $2.8168	\pm 2\times 10^{-4}$	& $1.1\times 10^3$\\
 & Err [$\%$] & $-4.48 \pm	0.35$ & $-5.35 \pm	0.32$\\
  \cmidrule(lr){2-6}
 & PBS-2 & $0.085533	\pm 6\times 10^{-6}$ & $2.9531	\pm 2\times 10^{-4}$	& $1.2\times 10^3$\\
 & Err [$\%$] & $-1.50 \pm	0.36$ &	$-0.77 \pm	0.34$\\
\midrule
3c & Poisson & $0.1974 \pm 7\times 10^{-4}$ & $8.15 \pm	0.02$ & $5.0 \times 10^{5}$\\
\cmidrule(lr){2-6}
 & Box & $0.1956	\pm 7\times 10^{-4}$ & $8.24	\pm 0.02$ & $4.6\times 10^5$\\
 \cmidrule(lr){2-6}
& CLS &  $0.183044 \pm 7\times 10^{-6}$ & $7.4839	\pm 10^{-4}$ & $1.1 \times 10^{4}$\\
& Err [$\%$] &  $-7.26 \pm	0.33$ & $-8.20 \pm	0.25$\\
\cmidrule(lr){2-6}
 & PBS-1 & $0.188079	\pm 7\times 10^{-6}$ & $7.7653	\pm 2\times 10^{-4}$ & $8.9\times 10^3$
\\
 & Err [$\%$] & $-4.71 \pm 0.34$ & $-4.75  \pm 0.26$\\
  \cmidrule(lr){2-6}
 & PBS-2 & $0.194130	\pm 8\times 10^{-6}$ & $8.0753	\pm 2\times 10^{-4}$	& $8.8\times 10^3$ \\
 & Err [$\%$] & $-1.64 \pm	0.36$ & $-0.95 \pm	0.27$\\
\bottomrule
\end{tabular}
\end{center}
\caption{Ensemble-averaged observables and computer time $t_{\mathrm{tot}}$ for the benchmark configurations: {\em suite} II - case $3$.\label{tab_suite2_case3}}
\end{table*}

\clearpage

\begin{figure*}
\begin{center}
\,\,\,\, Case 1a \,\,\,\,\\
\scalebox{0.7}{\input{flux_1A_I}}\,\,\,\,
\scalebox{0.7}{\input{flux_1A_II}}\\
\,\,\,\, Case 1b \,\,\,\,\\
\scalebox{0.7}{\input{flux_1B_I}}\,\,\,\,
\scalebox{0.7}{\input{flux_1B_II}}\\
\,\,\,\, Case 1c \,\,\,\,\\
\scalebox{0.7}{\input{flux_1C_I}}\,\,\,\,
\scalebox{0.7}{\input{flux_1C_II}}\\
\end{center}
\caption{Ensemble-averaged spatial scalar flux for the benchmark configurations: Case 1. Left column: {\em suite} I configurations; right column: {\em suite} II configurations. Solid lines represent the benchmark solutions obtained with the quenched disorder approach: green lines correspond to Markovian mixtures (Poisson tessellations), purple lines to Box tessellations. Dotted or dashed lines represent the solutions from annealed disorder approach: blue lines correspond to CLS results, black lines to PBS-1 results and red lines to PBS-2 results.\label{fig_space_1}}
\end{figure*}

\begin{figure*}
\begin{center}
\,\,\,\, Case 2a \,\,\,\,\\
\scalebox{0.7}{\input{flux_2A_I}}\,\,\,\,
\scalebox{0.7}{\input{flux_2A_II}}\\
\,\,\,\, Case 2b \,\,\,\,\\
\scalebox{0.7}{\input{flux_2B_I}}\,\,\,\,
\scalebox{0.7}{\input{flux_2B_II}}\\
\,\,\,\, Case 2c \,\,\,\,\\
\scalebox{0.7}{\input{flux_2C_I}}\,\,\,\,
\scalebox{0.7}{\input{flux_2C_II}}\\
\end{center}
\caption{Ensemble-averaged spatial scalar flux for the benchmark configurations: Case 2. Left column: {\em suite} I configurations; right column: {\em suite} II configurations. Solid lines represent the benchmark solutions obtained with the quenched disorder approach: green lines correspond to Markovian mixtures (Poisson tessellations), purple lines to Box tessellations. Dotted or dashed lines represent the solutions from annealed disorder approach: blue lines correspond to CLS results, black lines to PBS-1 results and red lines to PBS-2 results.
\label{fig_space_2}}
\end{figure*}

\begin{figure*}
\begin{center}
\,\,\,\, Case 3a \,\,\,\,\\
\scalebox{0.7}{\input{flux_3A_I}}\,\,\,\,
\scalebox{0.7}{\input{flux_3A_II}}\\
\,\,\,\, Case 3b \,\,\,\,\\
\scalebox{0.7}{\input{flux_3B_I}}\,\,\,\,
\scalebox{0.7}{\input{flux_3B_II}}\\
\,\,\,\, Case 3c \,\,\,\,\\
\scalebox{0.7}{\input{flux_3C_I}}\,\,\,\,
\scalebox{0.7}{\input{flux_3C_II}}\\
\end{center}
\caption{Ensemble-averaged spatial scalar flux for the benchmark configurations: Case 3. Left column: {\em suite} I configurations; right column: {\em suite} II configurations. Solid lines represent the benchmark solutions obtained with the quenched disorder approach: green lines correspond to Markovian mixtures (Poisson tessellations), purple lines to Box tessellations. Dotted or dashed lines represent the solutions from annealed disorder approach: blue lines correspond to CLS results, black lines to PBS-1 results and red lines to PBS-2 results.\label{fig_space_3}}
\end{figure*}

\end{document}